\documentclass[twoside]{articlek}

\usepackage[normal]{caption}

\renewcommand{\refname}{REFERENCES}

\def\BibTeX{{\rm B\kern-.05em{\sc i\kern-.025em b}\kern-.08em
		T\kern-.1667em\lower.7ex\hbox{E}\kern-.125emX}}

\usepackage{graphicx}

\renewcommand{\refname}{REFERENCES}

\def\BibTeX{{\rm B\kern-.05em{\sc i\kern-.025em b}\kern-.08em
		T\kern-.1667em\lower.7ex\hbox{E}\kern-.125emX}}

\usepackage{graphicx}

\begin{document}

\begin{flushleft}
	{\large\bf Evaporation Residue Production in Complete and Incomplete 
		Fusion Reactions of $^{20}$Ne with $^{197}$Au and $^{165}$Ho
		at E = 14.9 x AMeV
		}
	\vspace*{25pt}

	{\bf Fritz Peter He\ss berger$^{1,2,}\footnote{E-mail: \texttt{f.p.hessberger@gsi.de}}$} \\
	\vspace{5pt}
	{$^1$GSI - Helmholtzzentrum f\"ur Schwerionenforschung GmbH, Planckstra\ss e 1, 64291 Darmstadt, Germany\\
		$^2$Helmholtz Institut Mainz, Staudingerweg 18, 55128 Mainz, Germany}\\

\end{flushleft}
          
Version: April, 27, 2022\\

\abstract{
	Velocity distributions of heavy residues (A$_{res}$\,$>$\,A$_{tar}$), produced in irradiations of $^{197}$Au and $^{165}$Ho with $^{20}$Ne at
	bombarding energies of E = 208 MeV, have been measured at the velocity filter SHIP. While for the products from $^{20}$Ne + $^{165}$Ho just a mean value of
	v/v$_{CN}$ = 0.91$\pm$0.01 was obtained, individual residues, attributed to isotopes of astatine and polonium with mass numbers ranging from A\,=\,195 to 
	A\,=\,203 could be identified by means of $\alpha$ spectroscopy in the bombardments of $^{197}$Au. A linear increase of the mean velocity from 
	v/v$_{CN}$ $\approx$ 0.85 to v/v$_{CN}$ $\approx$ 0.93 with decreasing residue mass was observed. This effect was attributed to an increasing contribution 
	of pre-equilibrium particle emission to the production of heavy residues. A quantititative estimation resulted in a 90$\%$ contribution of pre-equilibrium processes 
	to the production of astatine isotopes, and a 50$\%$ contribution to the polonium isotopes.}

\section{1. Introduction}
At bombarding energies well above the fusion barrier, in competition to complete fusion, processes arise that are characterized by forming heavy composite nuclei close to the sum of the masses and nuclear charges of projectile and target, but having linear momenta smaller than those of the compound nuclei (CN),
having p$_{CN}$\,=\,p$_{proj}$. Since a connection between the missing mass and the missing linear momentum  is evident, such processes may be called 'incomplete fusion'. 
In principle, two scenarios are conceivable: firstly, a break-up of the projectile in the vicinity of the target nucleus, fusion of the latter one with one fragment, while the other one leaves the reaction zone without further interaction, as proposed in \cite{Siw79,Wil82}, or, secondly, emission of particles from the overlap region of the interacting nuclei in a very early stage of the reaction (see e.g. \cite{Blann81,Blann85}). In both cases no or only little interaction of the escaping particles with the target nucleus has occured, so their velocity is close to that of the projectiles and their angular distributions peak around the beam direction. Since they appear to be emitted before a complete amalgamation of projectiles and target nuclei has occured, i.e. thermal equilibrium was reached, they may be denoted as 'pre-equilibrium' particles. In this sense 'incomplete fusion' is accompanied by the emission of 'pre-equilibrium' particles, while vice versa the occurence of 'pre-equilibrium' particles is no unambiguous indication for incomplete fusion, however: the second projectile fragment may leave after, e.g. donating or recepting a few nucleons, and also projectile and target nuclei may reseparate again, without having formed a 'compound' nucleus.\\
In this sence we will distinguish here three cases, despite the fact, that naming may be treated in a different way in literature.\\
a) 'incomplete fusion': fusion-like processes without full momentum transfer, i.e. 'evaporation' of light particles with velocities close to the projectile velocities,
different to light particles emitted from a fully equilibrated composite nucleus ('compound nucleus') having a 'thermal' energy spectrum.\\
b) 'fusion after projectile break-up': break-up of the projectile in vicinity of the target nucleus; fusion of one projectile fragment with the target nucleus, while the other one escapes the reaction zone without further interaction.\\
c) 'transfer reactions': exchange of a 'couple' of nucleons between projectile and target nucleus at contact, followed by reseparation of both reaction partners.\\
In recent experiments we have investigated the production of heavy residues in bombardments of $^{197}$Au (at 114, 172 and 228 MeV) and $^{208}$Pb (at 172, 228 and 298 MeV) with $^{20}$Ne \cite{Hess91,Hess93,Hess94,Vel97}. Significant contributions of incomplete fusion to the heavy residue production, indicated by a shift of the mean velocity of the residues to values v$_{m}$/v$_{CN}$\,$\approx$\,0.90\,-\,0.95, was observed already at a bombarding energy of 172 MeV, which were interpreted as due to 'incomplete fusion'. In addition in reactions $^{20}$Ne + $^{208}$Pb a 'slow' residue component at v$_{m,slow}$/v$_{CN}$\,$\approx$\,(0.40-0.60) was observed, which appeared to decrease significantly towards the highest bombarding energy, in accordance with the model proposed in \cite{Wil82}, while the mean velocity of the 'fast' component was still decreasing.\\
As a completition of these measurements we here want to present the results for $^{20}$Ne + $^{197}$Au at E\,=\,208 bombarding energy.

\section{2. Experimental set-up}
The experiments essentially were performed with beams of $^{20}$Ne (E =\,11.4$\times$A MeV
(228 MeV) and 14.9$\times$A MeV (298 MeV)) 
from the UNILAC accelerator at GSI, Darmstadt. 
For light particle measurements also short irradiations at E = 5.9 AMeV (118 MeV) and 8.6 AMeV (172 MeV)  were 
performed.
The $^{20}$Ne beam had a repetition rate of 50 Hz and a duty factor of 23$\%$, i.e. each pulse of 4.6 ms duration was followed by a pause of 15.4 ms ('macro pulses').The 
$^{197}$Au targets were thin rolled foils of $\approx$180 $\mu$g/cm$^{2}$, while the $^{165}$Ho targets were produced by evaporating the material ($\approx$60 $\mu$g/cm$^{2}$)
on a carbon backing of $\approx$30 $\mu$g/cm$^{2}$ or were rolled foils of$\approx$600 $\mu$g/cm$^{2}$. Typical beam intensities of 30-50 pnA were used in the irradiations 
of $^{197}$Au and 5-10 pnA in the $^{165}$Ho irradiations.\\
The evapration residues (ER) recoiled from the targets nearly unretarded. They passed a thin ($\approx$15 $\mu$g/cm$^{2}$) carbon foil, which was placed 8 cm downstream the target position to equilibrate the ionic charge distributions of the evaporation residues. They were separated by the velocity filter SHIP \cite{Mun79} and were focussed to a detector system. Since in the current experiments a broad angular distribution was expected for the evaporation residues, the angular acceptance of SHIP was increased from 1.5$^{0}$ to 3.5$^{0}$ by placing the targets 80 cm closer to the entrance of the first quadrupole 
triplet. From model calculations \cite{Faust79,Reis85,Hess88} an enhancement of the efficiency by a factor 2--3 could be expected.\\
The detection system consisted of two transmission detectors \cite{Hess85}, 447 mm apart, that were used for the 'fine' velocity measurements of the evaporation residues, and two silicon detector arrangements, placed closely one after the other. The first one ('stop detector'), in which the evaporation residues were implanted, was a passivated ion implanted silicon detector of 60$\times$24 mm$^{2}$ and had an effective thickness of 300 $\mu$m. It was subdivided into ten strips, each of 6$\times$24 mm$^{2}$. Besides registration of the incoming evaporation residues, the 'stop detector' was used to measure the $\alpha$ - decay energies of the implanted nuclei. The energy resolution of the individual strips varied between (15--19) keV (FWHM) for $\alpha$ particles in the range (5--8) MeV.\\
The second detector ('light particle detector' (LPD)), an array of 3 $\times$ (24$\times$24) mm$^{2}$ passivated ion implanted silicon chips of 300 $\mu$m thicknesses, was used to detect high energetic light particles that passed SHIP and also the 'stop detector' and could pretent evaporation residues or $\alpha$ decays  due to their energy loss in the 'stop detector'. The combined system also allowed for light particle identification due to $\Delta$E$_{stop}$--E$_{LPD}$ or  $\Delta$E$_{stop}$--$\Delta$E$_{LPD}$ measurements.\\
For each registered event additionally the absolute time as well as the time within the 'macro pulse' were registered with continuously running clocks. These informations were used in the off-line analysis to distinguish between events occuring during the beam bursts and those occuring within the pause and also to measure decay curves after the end of each irradiation. The beam current was measured at the beam stop of SHIP. It was digitized and also registered.\\
The signals of the individual detectors were processed by conventional NIM/CAMAC modules and stored on tape event-by-event. Data aquisition and on-line analysis was performed with a VAX86 computer using the GOOSY system \cite{Essel88}, while off-line analysis was performed at an IBM3090-60J computer at GSI using the SATAN system \cite{Goeri88}.\\

\section{3. Efficiency cnsiderations}
To convert the measured particle rates into cross sections we have to consider the effciency of our set-up, which can be defined as the product of the SHIP transmission and the detection efficiency. The former is critically dependent on the angular distribution of the evaporation residues. Since for the products in question very small transmission values of one per cent are expected \cite{Faust79,Reis85}, already small variations, caused e.g. by deviation of the energy and angular distributions of the evaporated particles from those used in the model calculations, may lead to significant changes in the transmission values. As another consequence it follows that the SHIP transmission also depends on the evaporation channel. So, principally, a check of the calculated transmission values requires a complete set of independently measured cross sections for the same or a similar reaction, which is hardly feasible for the systems considered here. Thus we contended ourselves with a comparison of the evaporation residue cross section for $^{20}$Ne + 
$^{165}$Ho at E\,=\,290 MeV, where a value of $\sigma$\,=\,450 mb is reported \cite{Cramer89}.
(Here $\sigma$ is taken as the sum of evaporation residues from complete fusion and incomplete processes where one preequilibrium proton or one preequilibrium $\alpha$ - particle is emitted.) Using this value we obtain a SHIP transmission of $\epsilon$\,$\approx$\,0.034 averaged over all evaporation channels. To compare this experimental value with model calculations we used the fusion - evaporation codes HIVAP \cite{Reis81} as well as ALICE \cite{Blann82} to calculate the isotopic distribution of the residues, allowing as well emission of neutrons, protons and $\alpha$ particles as the emission of neutrons and  protons only. The sense of the latter procedure was to estimate the relative contributions of x$\alpha$ypzn - deexcitation channels and ypzn - deexcitation channels to specific residues. The cross sections for the individual residues were weighted with their calculated transmission values and integrated. As a result an averaged calculated transmission of $\epsilon$\,=\,0.027 was obtained. We are aware of the fact that preequilibrium emission may significantly change the isotopic distribution of the residues as well as the SHIP transmission for specific deexcitation channels. So the fair agreement between the experimental and calculated transmission may be somewhat fortitous.\\
For the somewhat heavier system $^{20}$Ne + $^{197}$Au we obtain from our calculations transmission values of $\epsilon$\,$\approx$\,(0.02-0.03) for ypzn - channels and
$\epsilon$\,$\approx$\,(0.005-0.001) for x$\alpha$ypzn - channels.\\
Conservatively we do not claim the cross section values more reliable than within a factor of three.\\

\begin{figure*}
\resizebox{0.99\textwidth}{!}{%
	\includegraphics{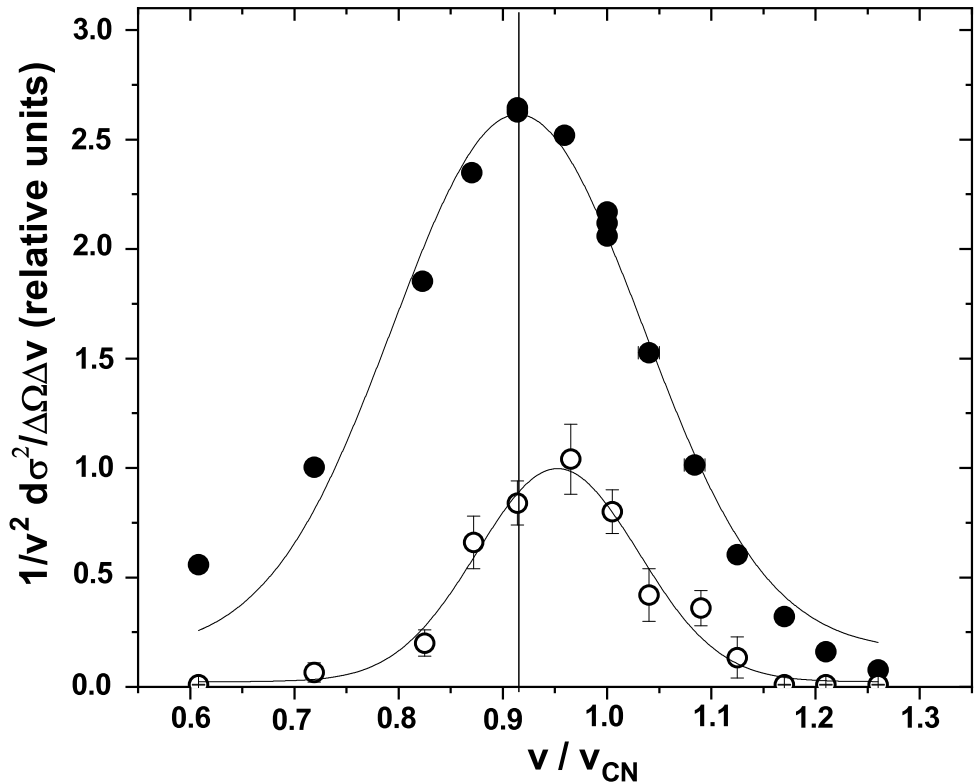}
}
\caption{velocity distributions for evaporation residues from  $^{20}$Ne + $^{165}$Ho at E = 298 MeV. Full dots: all evaporation residues; open circlers: $^{170}$Os}
\label{fig:1}       
\end{figure*}

\begin{figure*}
\vspace{-1cm}   
\resizebox{0.99\textwidth}{!}{%
	\includegraphics{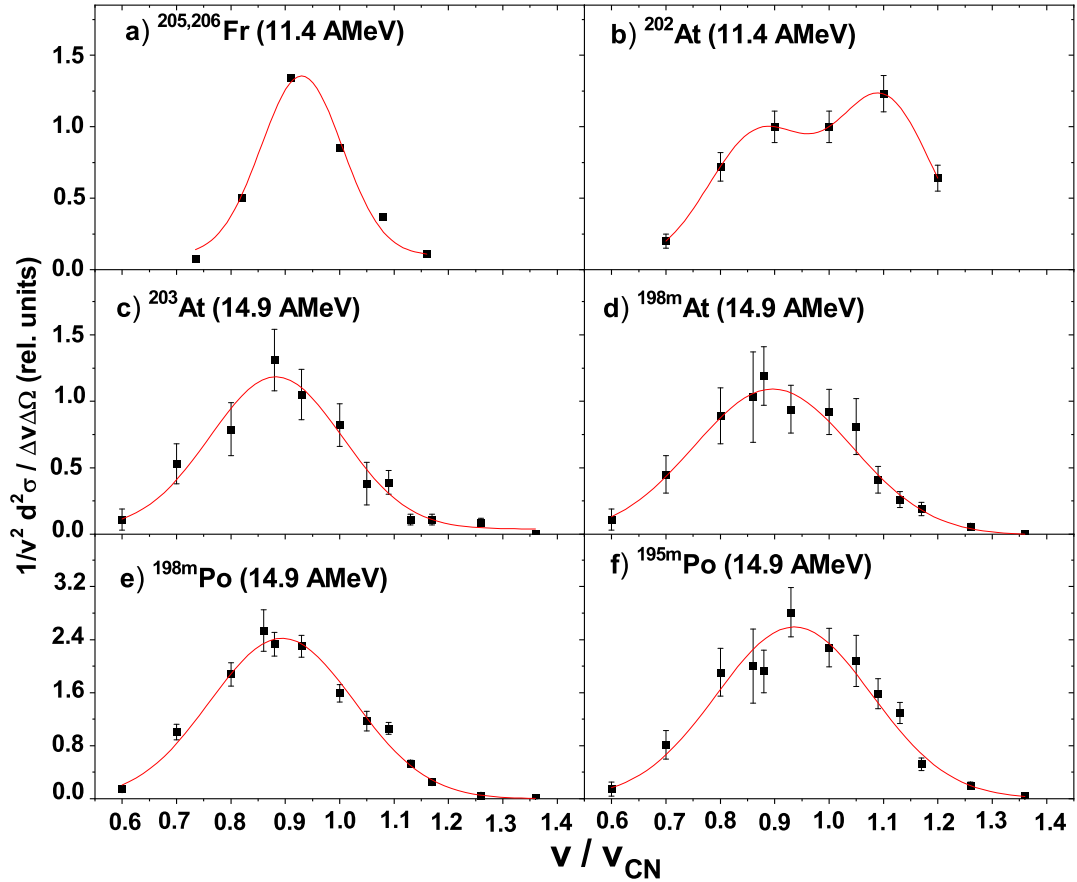}
}
\caption{velocity distributions for evaporation residues from  $^{20}$Ne + $^{197}$Au}
\label{fig:2}       
\end{figure*}

\vspace{10mm}

\section{4. Experimental results}
\subsection{\bf{4.1 Velocity distributions}}
Since SHIP is a Wien type velocity filter, velocity distributions of evaporation residues can be measured just by changing the E/B - ratio (see ref.\cite{Hess91} for details). A drawback of this method, however, is that the velocity distribution of the evaporation residues is folded with the (velocity dependent) transmission of SHIP. As a consequence, the mean velocity of the detected residues in general will be slightly different from the preset velocity defined by the E/B - ratio. To overcome this problem we measured the velocities explicitly using the time-of-flight system mentioned above. The velocities obtained were corrected for the energy loss in the carbon foil ($\approx$ 30 $\mu$g/cm$^{2}$) of the first transmission detector using the energy loss table of ref. \cite{North70}. The deviation between the preset velocity v$_{E/B}$ and the measured velocity v$_{TOF}$ was found to be $|$1-v$_{E/B}$/v$_{TOF}$$|$$\le$0.03.\\
The velocity distributions measured for $^{20}$Ne + $^{165}$Ho are shown in fig. 1. In this reaction only one $\alpha$ emitter, $^{170}$Os (open circles) which represents the p14n - channel of the compound nucleus $^{185}$Ir was registered with an intensity high enough to evaluate a velocity distribution. Its production rate was only
a fraction of 2.4$\times$10$^{-4}$ of the total evaporation residue production rate.\\
The full dots in fig. 1 represent the distribution for the evaporation residues, registered by ion counting. While the distribution of $^{170}$Os peaks at 
v/v$_{CN}$\,=\,0.95$\pm$0.01, that for the evaporation residues peaks at a significantly lower value of v/v$_{CN}$\,=\,0.91$\pm$0.01. The difference can be 
interpreted due to a different number of particles emitted predominantly in forward direction, denoted in the following  as 'preequilibrium particles' despite of possibly originating from different nuclear processes. For $^{170}$Os the emission of one preequilibrium particle (proton or neutron) on the average is suggested. For the evaporation residues, as they represent the sum over all deexcitation channels, weighted by the SHIP transmission (see sect. 3) the situation is more complicated. As the bulk of the mass distribution is represented by isotopes which are no $\alpha$ emitters, most probably those with Z\,$<$\,76 and N\,$>$\,94, the value of v$_{m}$/v$_{CN}$ just represents a mean value, that does not allow for a definite conclusion about the mass or the number of emitted preequilibrium particles in each process leading to an evaporation residue with a definite A$_{ER}$ and Z$_{ER}$. It just may be concluded that it represents the bulk of residues where up to four mass units were taken away by preequilibrium particles.\\
A significant deviation from the shape of a single Gaussian, however, is evident for evaporation residues at 
v$_{m}$/v$_{CN}$\,$<$\,0.7.This behavior is interpreted as due to contributions of incomplete fusion fractions with mass transfer A$_{tr}$\,$<$\,16, or, vice versa, the removement of four units of mass by preequilibrium emission, in accordance with the results in ref. \cite{Cramer89}.\\  
Typical examples of velocity distributions for evaporation residues from irradiations of $^{197}$Au with $^{20}$Ne are shown in figs. 2c-f. For comparison we present examples from the 11.4 AMeV - measurement in a previous experiment \cite{Hess91} in figs. 2a, 2b. Different to $^{202}$At produced at 11.4 AMeV (fig. 2b) for none of the produced isotopes an enhanced intensity at v/v$_{CN}$\,$>$\,1 is clearlv indicated at 14.9 AMeV, i.e. contributions from emission of thermal $\alpha$ particles to the evaporation residue production must be regarded as negligible.\\
Instead we observe, similar to $^{205,206}$Fr produced at 11.4 AMeV rather symmetric distributions with maxima shifted to v$_{m}$/v$_{CN}$\,=\,(0.86-0.94).\\
The positions of the maxima were found to depend significantly from the residues mass A$_{ER}$: we unambiguously observe the highest mean velocities for the lightest residues. A dependence of v$_{m}$/v$_{CN}$ from the atomic number Z$_{ER}$ (fig.3) of the residues is not indicated.
The mass dependence v$_{m}$/v$_{CN}$ can be understood as a consequence of an interplay between emission of 'thermal' and preequilibrium particles during deexcitation. 
Since the velocities (and thus the energies) of the preequilibrium particles are higher than those of thermal particles, they will cool down the excited nucleus more effective than thermal ones. Preequilibrium emission thus will lead to heavier residues. Since the angular distributions of preequilibrium particles is peaked in forward direction, while thermal particles are emitted essentially isotropic, a net shift to lower mean velocities will occur. The magnitude of the shift will be governed by the multiplicity, mass and energy of the preequilibrium particles emitted in the process leading to a specific residue.
The widths of the distributions (FWHM) vary between 24$\%$ and 34$\%$ of v$_{CN}$. An unambiguous dependence on the residue masses, however, is not indicated.\\ 

\begin{figure*}
\resizebox{0.99\textwidth}{!}{%
	\includegraphics{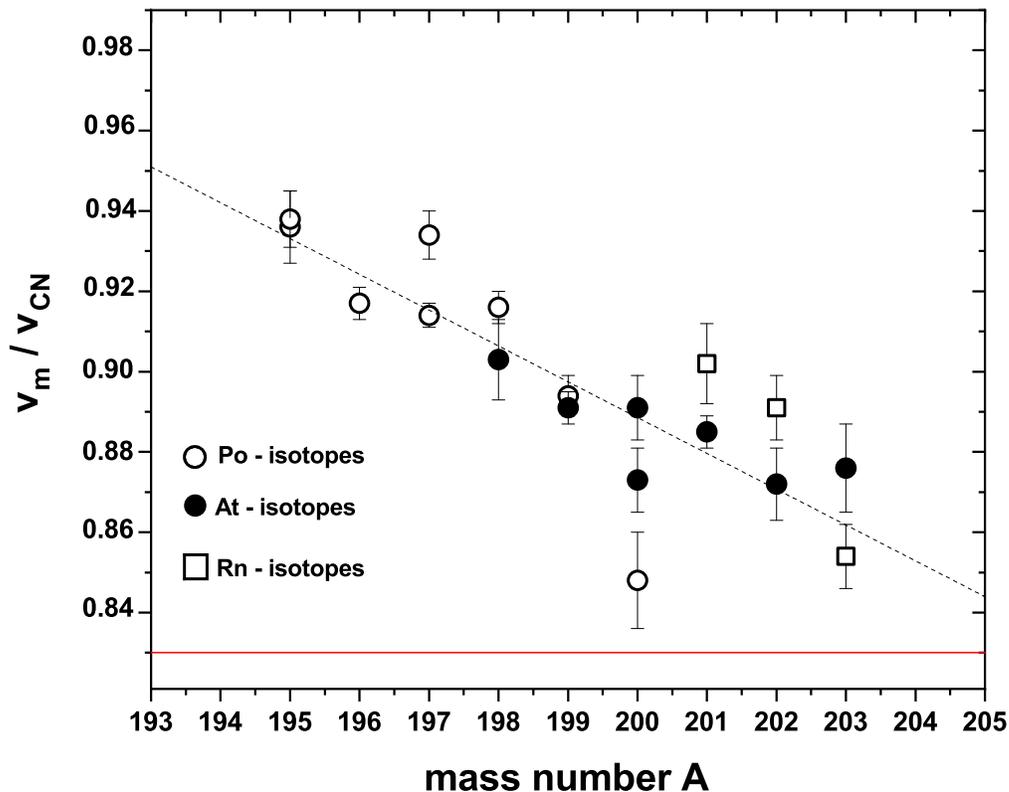}
}
\caption{mean velocities of evaporation residues from  $^{20}$Ne + $^{197}$Au at E = 298 MeV; open circles = polonium isotopes, full circles = astatine isotopes, 
	open squares = radon isotopes}
\label{fig:3}       
\end{figure*}

\subsection{\bf{4.2 Cross sections}}
Cross sections can be deduced from the intensity of the $\alpha$ lines according to the efficiency considerations in sect. 2.\\
From the $\alpha$ spectra we obtain total cross sections of $\sigma$(At)\,=\,825$\pm$63 $\mu$b for the At-isotopes. For the Po - isotopes we observe a comparable cross-section of
$\sigma$(Po)\,=\,572$\pm$20 $\mu$b, while for the Rn - isotopes a contribution of only about 10 $\mu$b is obtained. For the total amount of evaporation residues, registered by ion counting, we obtain a value $\sigma$(ER)\,=\,2.9$\pm$0.1 mb. The $\alpha$ emitters thus contribute to about 48 $\%$ of the total residue cross-section. The cross-sections for the individual residues, identified by their $\alpha$ decay, are displayed in fig. 4 in comparison with model calculations (see sect. 5.2). For astatine we obsereve the maximum cross-section at about A\,=\,200, while for polonium the values appear almost constant in the region A\,=\,196 and A\,=\,200. The bulk
of the residues ($\approx$52$\%$) seems to be represented by Po - isotopes with masses A\,$>$\,200 and nuclei with atomic numbers Z\,$<$\,84, which have $\alpha$ branches b$_{\alpha}$\,$<$\,0.05 and are not identified unambiguously in our experiment. The shift by about two mass units from Z\,=\,85 to Z\,=\,84 may indicate that the polonium isotopes are produced to a higher extent by emission of preequilibrium $\alpha$ particles from the composite nucleus.\\ 

\begin{figure*}
	\vspace{-1.2cm}
	\resizebox{1.00\textwidth}{!}{%
		\includegraphics{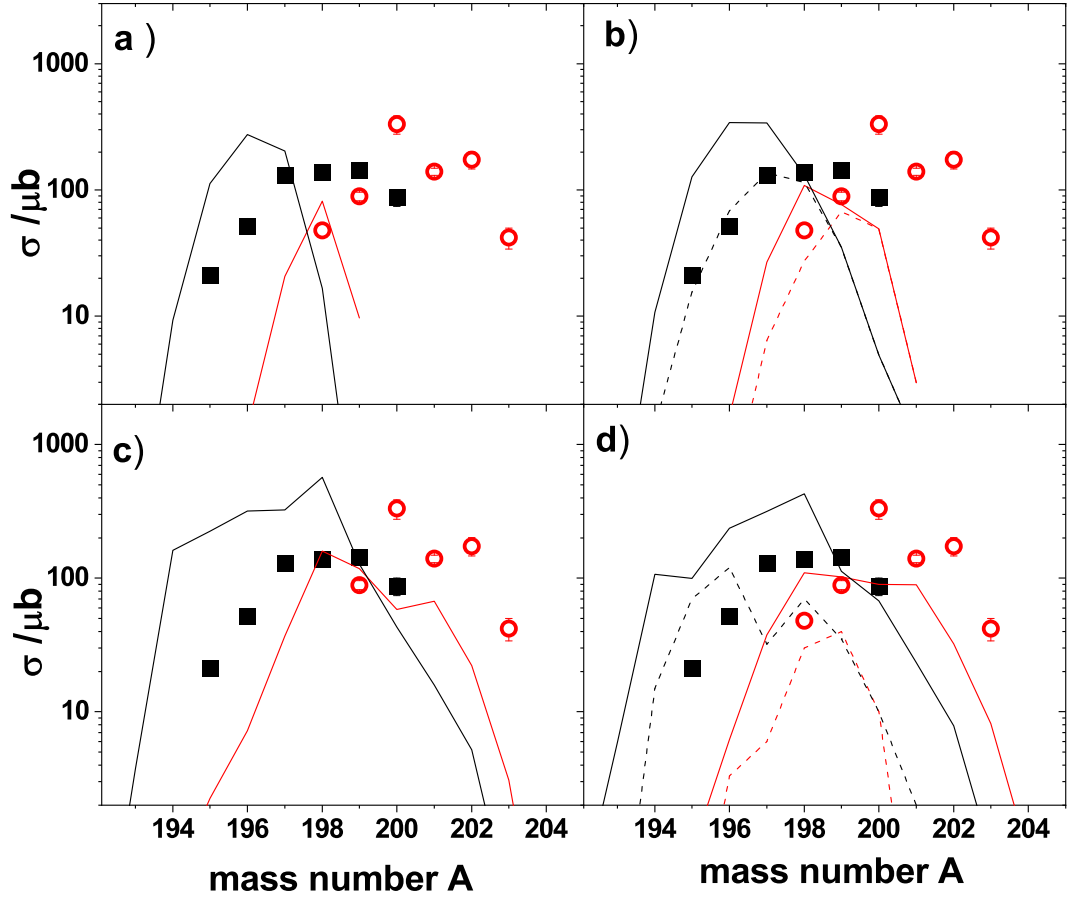}
	}
	\caption{evaporation residue cross sections for  $^{20}$Ne + $^{197}$Au at E = 298 MeV, full squares: polonium isotopes, open circles: astatine isotopes.
		The lines are the result of calculations for Po- isotopes (black) and At- isotopes (red); a) full lines: HIVAP calculations, complete fusion; b) full lines: 
		sum rule model + HIVAP, complete and incomplete fusion; dashed lines: sum rule model + HIVAP, only incomplete fusion; c) full lines: ALICE calculations; d) full lines: ALICE  + standard hybrid model, dashed lines:
		sum rule model + ALICE. }
	\label{fig:4}       
\end{figure*}

\vspace{5mm}
\subsection{\bf{4.3 Light particle measurements}}
Due to their high magnetic and electric rigidity high energetic light particles (protons, $\alpha$ particles etc.) are only weakly deflected by SHIP; thus 
they can pass the separator and be detcted. For this purpose two Si-detectors, forming a 'telescope' were mounted in the focal plane of SHIP. Despite the small
solid angle ($\Omega$ $\approx$1.2 $\times$ 10$^{-6}$) due to their high production cross sections notable amounts of such particles can be registered in relatively
short irradiation times. To ensure that indeed the particles stem from reactions with the target material, such measurements were performed only with the gold targets,
as they did not have carbon backings. Also the carbon converter foil was removed. The measurements were performed at four bombarding energies  E$_{lab}$ =
118, 172, 228 and 298 MeV. A two-dimensional plot of energies or respectively energy losses $\Delta$E$_{1}$ ('stop detector') versus $\Delta$E$_{2}$ 
('light particle detector') at the bombarding energy of E$_{p}$\,=\,298 MeV is shown in Fig. 5. Four groups of particles could be identified: a)  Z\,=\,1  - particles, protons or deuterons (1),
b) Z\,=\,2 - particles, $\alpha$ particles or $^{3}$He nuclei (2), c) Z\,=\,3 - particles, Li - nuclei (3), and d) Z\,=\,4 - particles, Be-nuclei (4). 
This arrangement allowed for detection of, e.g. $\alpha$ - particles in the range E $\approx$ 22--32 MeV. $\alpha$ - particles of lower energy (E$<$23 MeV)
are fully absorbed in the stop detector and thus are not visible in the $\Delta$E$_{1}$ -- $\Delta$E$_{2}$ plane. At higher particle energies $\Delta$E$_{1}$ decreases with increasing 
energy while the energy $\Delta$E$_{2}$ measured in the LPD increases (upper branch) until a value $\Delta$E$_{2}$\,$\approx$\,25 MeV is reached. Particles 
of higher energies can punch through the LPD, and we observe both, decreasing $\Delta$E$_{1}$ and decreasing $\Delta$E$_{2}$ values. The bulk of the distribution was found at 
$\Delta$E$_{1}$ $\approx$ $\Delta$E$_{2}$ $\approx$ 7.5 MeV, corresponding to a total of E\,=\,55 MeV, reconstructed using the energy loss of $\alpha$ particles
in silicon according to \cite{HuB90}. The energy distributions of the $\alpha$ particles could be divided into two components: a) one of E\,$\approx$\,30 MeV, only
varying slightly with the beam energy, which was ascribed to 'thermal - evaporation' $\alpha$ particles, and b) a component with energies suggesting velocities close to the
projectile velocity (shown in fig. 6), which was ascribed to 'preequilibrium' $\alpha$ particles.\\
It should be remarked, that at the specific beam energies neither the energy nor the intensity of the preequilibrium $\alpha$ - particles depended on a specific
velocity setting of SHIP, which means, that these particles past SHIP nearly undeflected.

\begin{figure*}
		\vspace{-1cm}   
	\resizebox{0.99\textwidth}{!}{%
		\includegraphics{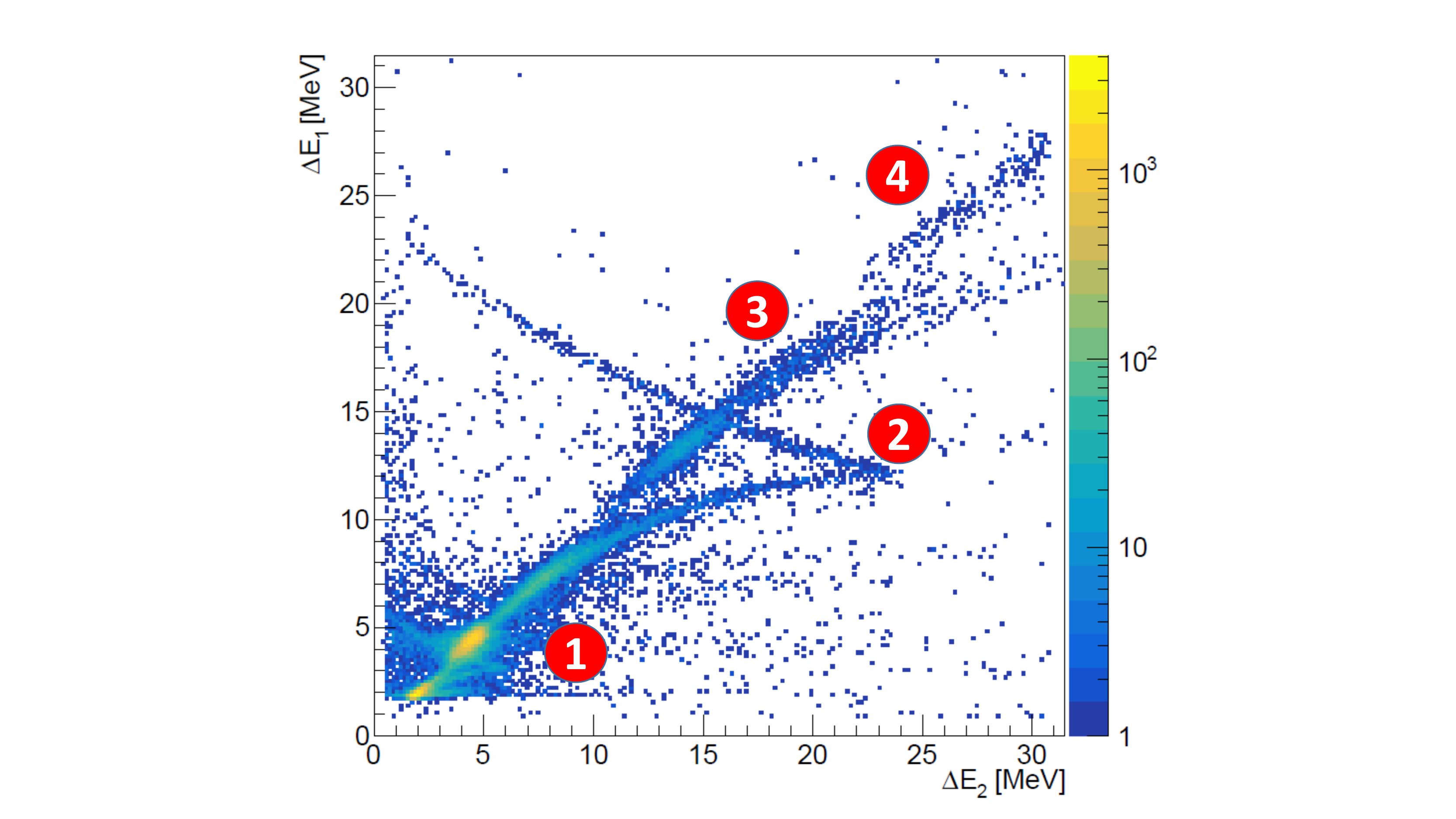}
	}
	\caption{Light particle spectra measured with the detector telescope
		('stop detector'  + 'light particle detector' (LPD)) behind SHIP.
	}
	\label{fig:5}       
\end{figure*}

\begin{figure*}
	\vspace{-2.1cm} 
	\resizebox{0.99\textwidth}{!}{%
		\includegraphics{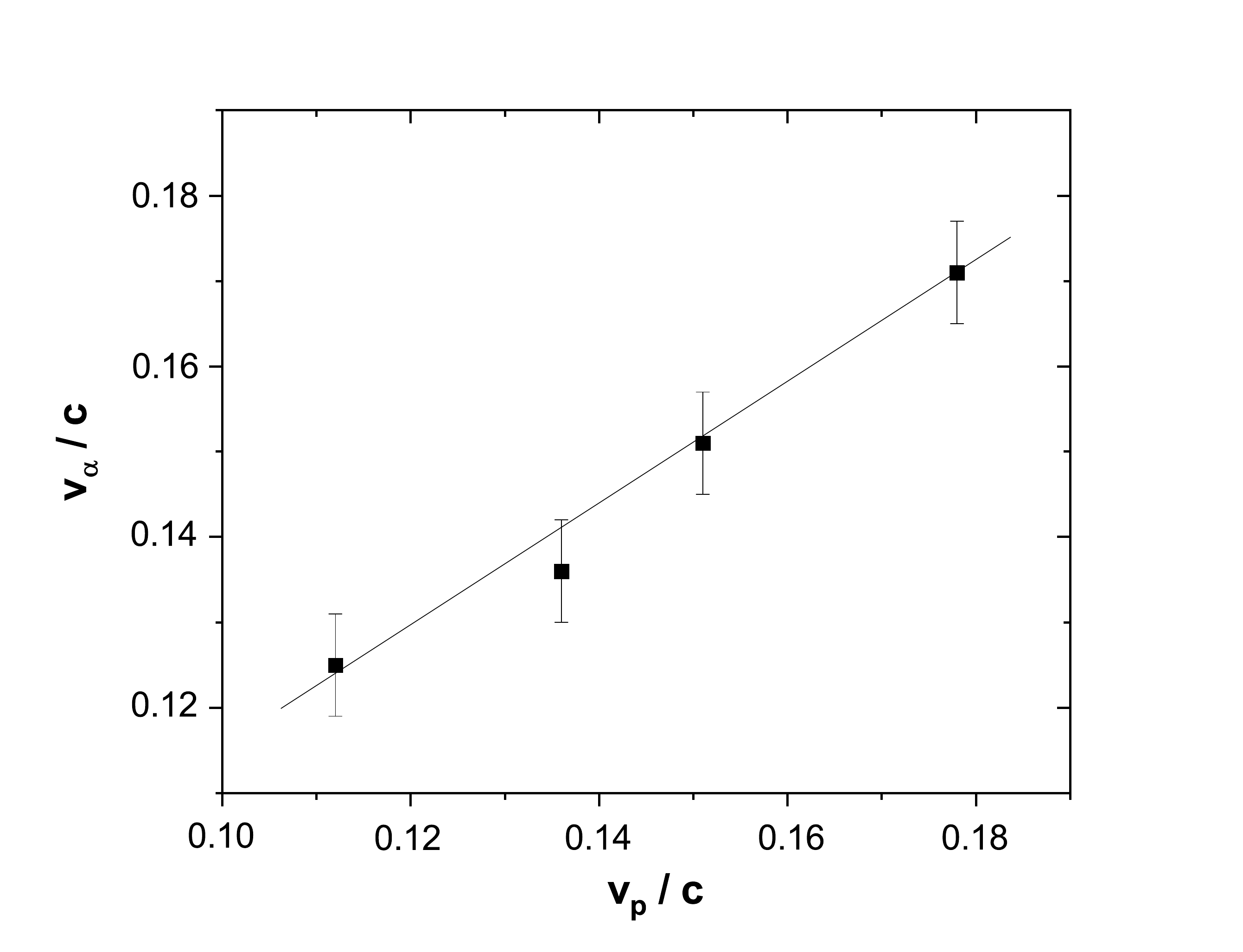}
	}
	\caption{Mean velocity (v$_{\alpha}$) of the 'high enery' $\alpha$ particle
		component as function of the projectile velocity (v$_{p}$)
	}
	\label{fig:6}       
\end{figure*}

\begin{figure*}
	\vspace{-2.1cm}
	\resizebox{0.99\textwidth}{!}{%
		\includegraphics{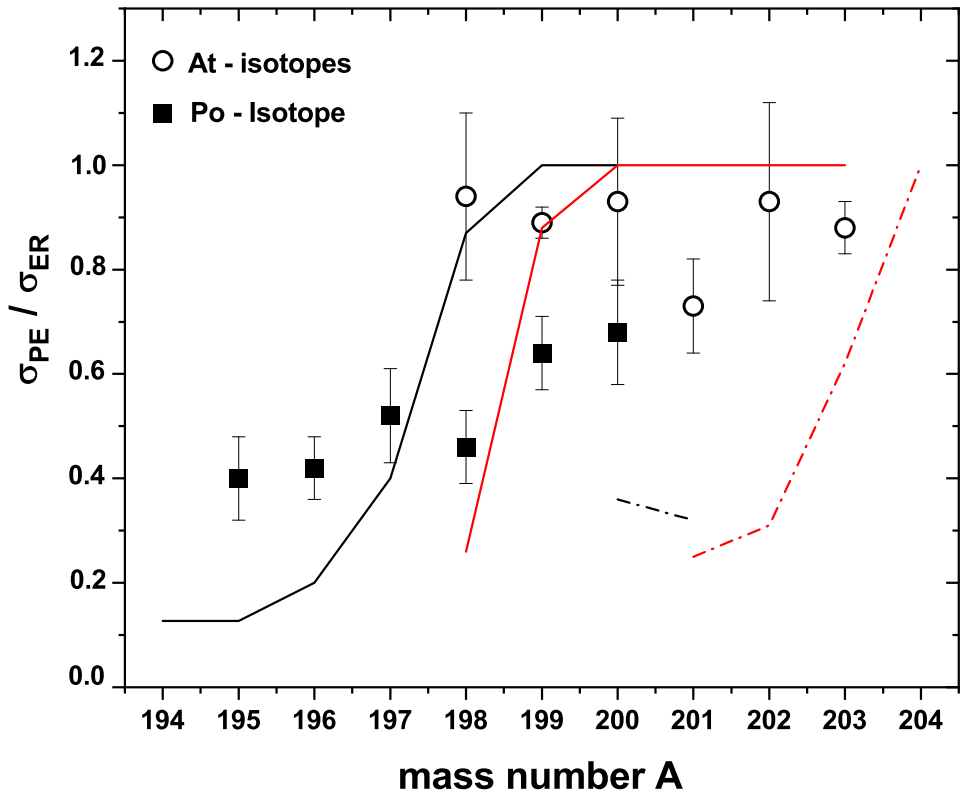}
	}
	\caption{Fraction of residues from incomplete fusion to total. Open circles: astatine isotopes, full squares: polonium isotopes; full lines: 
		results from calculations using 
		sum rule + HIVAP for polonium (black) and  astatine - isotopes (red); dashed dotted lines: results from ALICE + standard hybrid model calculations
	}
	\label{fig:7}       
\end{figure*}

\begin{figure*}
	\vspace{-2.1cm}
	\resizebox{0.99\textwidth}{!}{%
		\includegraphics{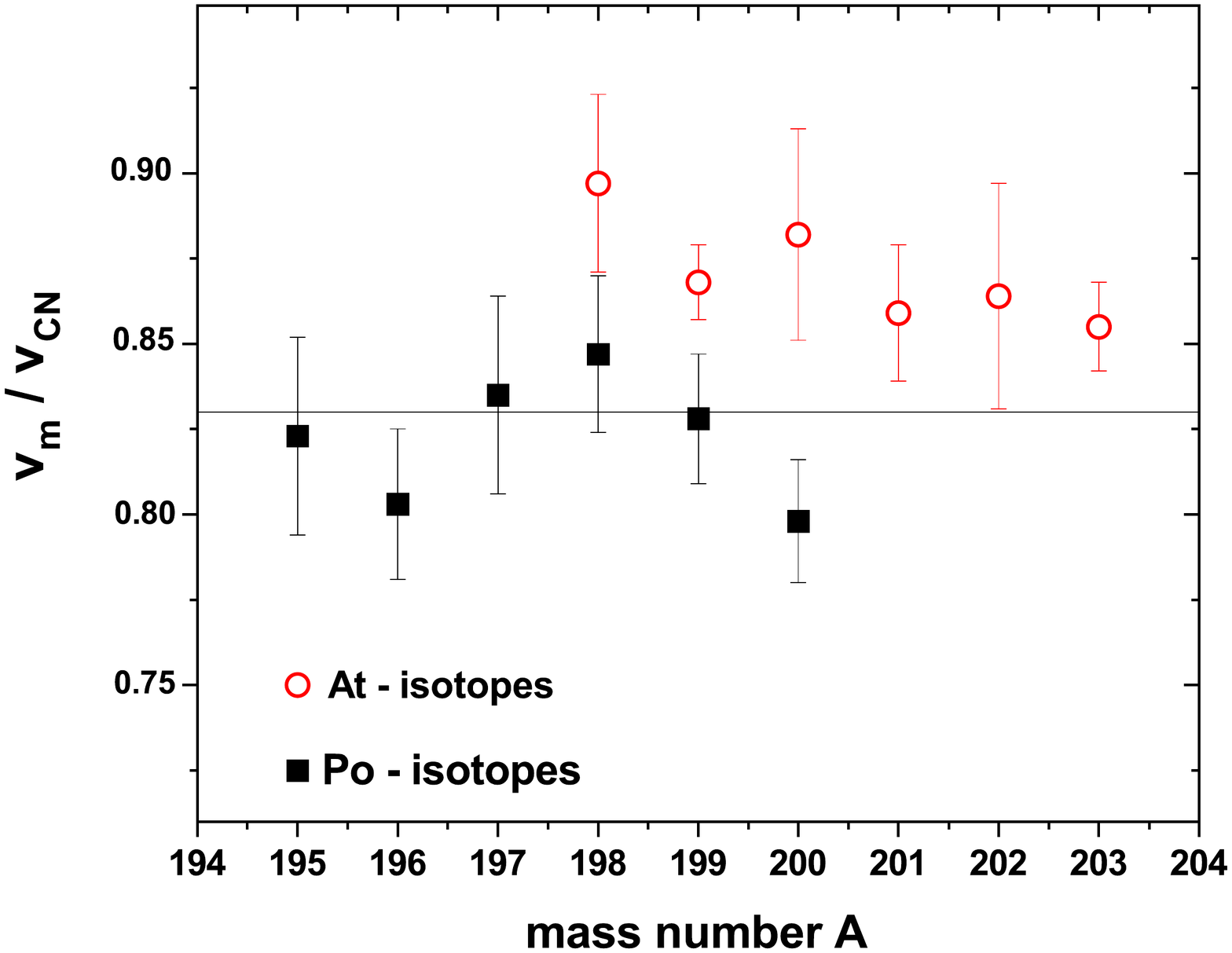}
	}
	\caption{mean velocities of the incomplete fusion component of evaporation residues from  $^{20}$Ne + $^{197}$Au at E = 298 MeV; open circles: astatine isotopes, full squares: polonium isotopes. The line represents the mean velocity obtained from fusion-fission reactions.}
	\label{fig:8}       
\end{figure*}

\vspace{10mm}

\section{5. Discussion}
\subsection{\bf{5.1 Incomplete Fusion}}
As it was already discussed in sect. 4.1, the velocity distributions for the different residues reflect a competition between emission of only thermal particles and, in addition, also preequilibrium particles during the deexcitation of the hot composite nuclei.\\
Thus it seems meanigful to decomposite the measured velocity distributions into a 'complete' fusion (emission of only thermal particles) and an 'incomplete' fusion (emission of preequilibrium particles in addition to thermal particles) fraction. The velocity distribution for complete fusion was estimated by model calculations
\cite{Faust79,Reis85}, taking into account emission of thermal nucleons (protons and neutrons) only. To determine both fractions we fitted the measured velocity distributions  each by two gaussians. Peak positions and FWHM for that representing complete fusion were taken from the model calculations. Strictly, one would have also to respect emission of thermal $\alpha$ particles, which, however leads to strong components in the distributions for v/v$_{CN}$\,$>$\,1. Experimentally such contributions are not evident for $^{20}$Ne + $^{197}$Au at E\,=\,14.9 AMeV (see figs. 2c-2f). Fits including also velocity distributions expected for $\alpha$ - channels in the complete fusion fraction resulted in contributions of $<$0.01 for the At-isotopes and $<$0.05 for the Po-isotopes. Thus contributions from $\alpha$xpyn channels were neglected. The ratio of the peak areas $\gamma$\,=\,A$_{inc}$/(A$_{inc}$\,+\,A$_{com}$), where A$_{inc}$, A$_{com}$ denote the areas of the gaussians attributed to incomplete and complete fusion, respectively, was taken as a measure for the completeness  of the fusion leading to the production of the specified evaporation residue. The results are shown in fig. 7. A significantly different behavior for the At (Z=85) and Po (Z=84) - residues is evident. The 'incomplete fusion' fractions for the At-isotope is constant, typically around 90$\%$ of the total, while for the Po-isotopes it is increasing from $\approx$40$\%$ to $\approx$70$\%$ with increasing residue mass number. Qualitatively this behavior might be understood as due to the influence of different multiplicities and masses of preequilibrium particles.
This behavior, however, is not well reproduced by model calculations using either the sum rule model for calculating the reaction cross sections and HIVAP for deexcitation (full lines in fig.7) or ALICE and the standard Hybrid model (dashed-dotted lines in Fig.7). While stable values around  $\gamma$\,=\,1 are obtained for A\,$\ge$\,199
(Po - isotopes) and  A\,$\ge$\,200 (At - isotopes) and a steep decrease towards lower mass values are obtained from the calculations, stable values of
$\gamma$\,$\approx$0.90 for the At-isotopes and a moderate increase of $\gamma$ with increasing mass number are observed for the Po-isotopes. The ALICE + standard hybrid model calculations (dashed-dotted lines), however, do not reproduce the experimental data at all.\\ 
A somewhat different behavior is found for the maxima (or 'mean velocities') of the 'incomplete fusion' component as obtained from the fits and shown in fig. 8. Despite of large error bars and straggling of the data we observe significantly different values for the At- and Po-isotopes, We obtain mean values, averaged over all isotopes observed, of (v/v$_{CN}$)$_{incomp}$\,=\,0.87$\pm$0.02 for the astatine and  (v/v$_{CN}$)$_{incomp}$\,=\,0.82$\pm$0.02 for the polonium isotopes. For the At-isotopes further a trend to decreasing mean velocities for increasing masses is indicated. These values indicate that in incomplete fusion processes leading to the production of At-isotopes about (2-3) nucleons are emitted on the average prior to thermal equilibrium, in those leading to Po-isotopes four mass units are carried away on the average (probably predominantly as an $\alpha$ particle) by preequilibrium processes. Thus the Po-isotopes are produced with nearly equal fractions either by incomplete fusion, accompanied by emission of preequilibrium particles, or by complete fusion followed by emission of thermal nucleons, but only to a small extent by emission of thermal $\alpha$ particles during deexcitation. Contrary to the situation at lower bombarding energies \cite{Hess91}, where we observed a competition between preequilibrium and thermal $\alpha$ particles, we seemimgly here have an 'either - nor' situation, i.e. either a preequilibrium $\alpha$ particle is emitted or no $\alpha$ particle will be emitted during the deexcitation process.\\
Our experiments indicate higher mean velocities of the evaporation residues than the value obtained from linear momentum transfer measured in fusion-fission reactions \cite{Tubbs85} assuming the missing momentum is carried away by one or several particles moving in beam direction and with the projectile velocity (see fig. 3). Comparing the results of \cite{Tubbs85} with the results of the incomplete component for the individual isotopes (see fig. 8), an agreement between fusion - fission results and the velocities for the Po-isotopes is indicated, while those for the At-isotopes the mean velocity is higher. In \cite{Hess91} we compared the intensity ratio of preequilibrium and thermal $\alpha$ particles measured in \cite{Fuchs85} and our cross-section ratios for evaporation residue production including preequilibrium emission and solely thermal emission. We found that only a small part of the preequilibrium $\alpha$ particles contributed to the evaporation residue production (compared to thermal ones) and concluded that preequilibrium production was connected to high angular momenta.\\  
This interpretation is in-line with the sum-rule model of Wilczinski et al.\cite{Wil82}; a) break-up of $^{20}$Ne into $^{16}$O + $^{4}$He followed by fusion of $^{16}$O with $^{197}$Au, while the $\alpha$ particle moves on with beam energy and is considered as a 'preequilibrium particle'. This process is limited to a narrow l-window close to the grazing angular momentum, leaving the incomplete compound nucleus $^{213}$Fr at a high angular momentum of $\hbar$\,$\approx$\,68, where its fission barrier is zero, and an excitation energy of E$^{*}$\,$\approx$\,130 MeV; so at a bombarding energy of 11.4 AMeV only branches of low l-values can contribute to the evaporation resudue production.
The same holds for break-up $^{20}$Ne $\rightarrow$ $^{19}$F + p etc.. At E = 14.9 AMeV we expect from the sum-rule model about the same angular momentum, but a higer excitation energy of  E$^{*}$\,$\approx$\,180 MeV; b) in bombardments of $^{208}$Pb with $^{20}$Ne 'slow' evaporation residues were attributed to (incomplete) fusion after projectile break-up into more symmetric fractions, leading to lower angular momenta of the composite system were observed \cite{Hess94}. This component was present at bombarding energies E = 8.6 AMeV and 11.4 AMeV, but not observed at E = 14.9 AMeV.\\

\subsection{\bf{5.2 Mass Distributions}}
To achieve further insight we compared our measured evaporation resuidue cross sections with model calculations.  In one set of calculations we used the sum-rule model to obtain the reaction cross-section as well as the excitation energies and angular momentum distributions for the products from complete fusion reactions and the different incomplete fusion channels, and used the HIVAP code \cite{Reis81} for deexcitation. Alternatively we used the ALICE code \cite{Blann82}. The results are presented in fig 4.\\
In figs. 4a, 4b we present the results using the sum-rule model plus HIVAP. According to recent results on 'fission hindrance' (see e.g. \cite{Hilsch92}) at high excitation energies, the $\Gamma_{f}$ - value from HIVAP was scaled by a parameter exp((E$^{*}$-E$_{th}$/E$_{\tau}$)). Using E$_{th}$\,=\,90 MeV and E$_{\tau}$\,=\,20 MeV (we do not stress to much attention to these specific numbers), a reproduction of the residue cross section values by a factor of five was achieved in a narrow region of residue masses. But neither respecting complete fusion alone (fig. 4a), nor including incomplete fusion component (fig. 4b) by summing all contributions from projectile fragments A$_{pf}$ $\le$ A$_{p}$ - 6 
(lighter projectile fragments lead to residues with lower velocities) were able to reproduce the measured mass distributions for both, the At- and the Po-isotopes. The calculated mass distributions are narrower and shifted towards lower masses. Even respecting incomplete fusion, we still fail to reproduce the cross sections for the heaviest masses.\\
A similar situation is evident for the ALICE - calculations. We get here a broader mass distribution already when respecting only complete fusion (fig. 4c). The inclusion of preequilibrium emission using the hybrid model  with standard parameters  results in a notable increase of the cross-sections on the heavy mass side, but still underpredicts the experimental data (fig 4d). A similar situation is observed when the results from the sum-rule model calculations are used as input for deexcitation calculations with ALICE (dashed lines in fig. 4d). Here the steep decrease sets in already at lower mass numbers.\\
It is further evident, that in all types of calculations the cross sections for the lighter polonium isotopes are overestimated in general, while those for the heavier ones are underestimated. For the astatine isotopes they are underestimated in general. Although, in prinviple, it cannot be excluded, that this behavior is due to an incorrect treatment of the fission barriers, we suppose a higher contribution of incomplete fusion than predicted by the models used here, leading to an enhanced production of heavier residues. It seems, however, questionable that this fraction can be attributed to a higher contribution from fusion after projectile break-up rather than by 
pre-equilibrium emission, as this reaction type would lead to lighter residues, as shown by both the HIVAP and the ALICE results.\\
So tentatively we would thus ascribe the enhanced cross sections for the heavier masses to particle emission from the interaction zone in an early stage of the reaction, as described, e.g. in \cite{Blann81,Blann85}, but with higher emission rates and probably with higher multiplicities or higher masses. The angular momentum distribution of composite systems produced in such processes were recently investigated theoretically by Krishan et al. \cite{Krishan92}. They found a considerable lowering of the angular momentum compared to complete fusion, e.g. by 20$\hbar$ for $^{27}$Al + $^{84}$Kr at E = 20 AMeV.\\
Since emission of preequilibrium particles further reduces the excitation energy, the influence of both further could contribute to an enhancement of the stability against prompt fission and thus to enhanced residue cross-sections.\\

\begin{figure*}
	\vspace{-25mm}
	\resizebox{0.90\textwidth}{!}{%
		\includegraphics{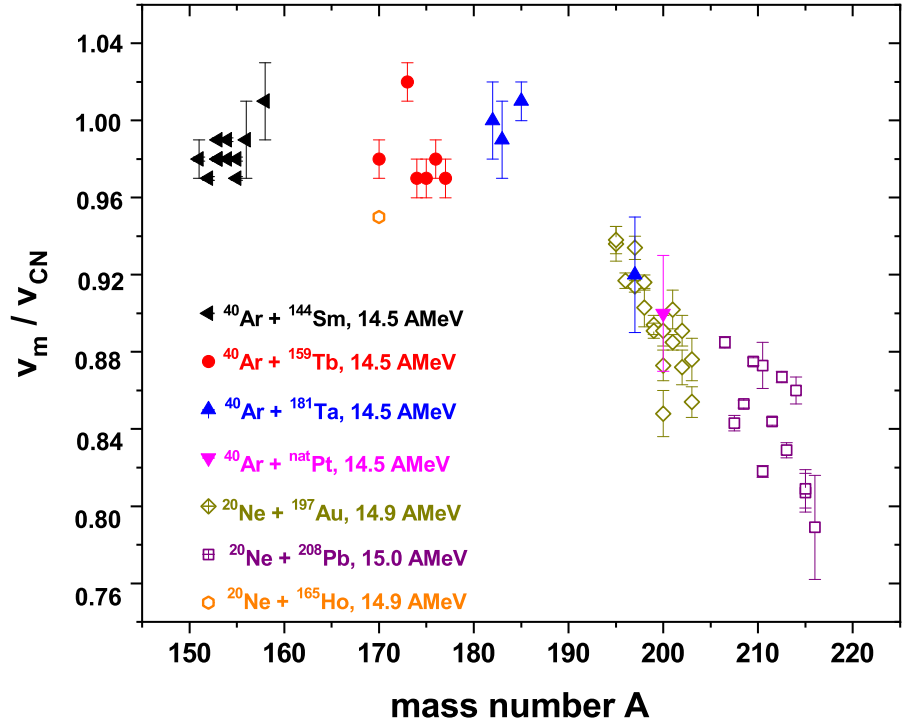}
	}
	\caption{Mean velocities of evaporation residues from  $^{40}$Ar + $^{144}$Sm,$^{159}$Tb, $^{181}$Ta, $^{nat}$Pt at E = 580 MeV,  $^{20}$Ne + $^{165}$Ho, $^{197}$Au at E = 298 MeV, $^{20}$Ne + $^{208}$Pb at E = 300 MeV. }
	\label{fig:9}       
\end{figure*}

\begin{figure*}
	\vspace{-20mm}
	\resizebox{0.99\textwidth}{!}{%
		\includegraphics{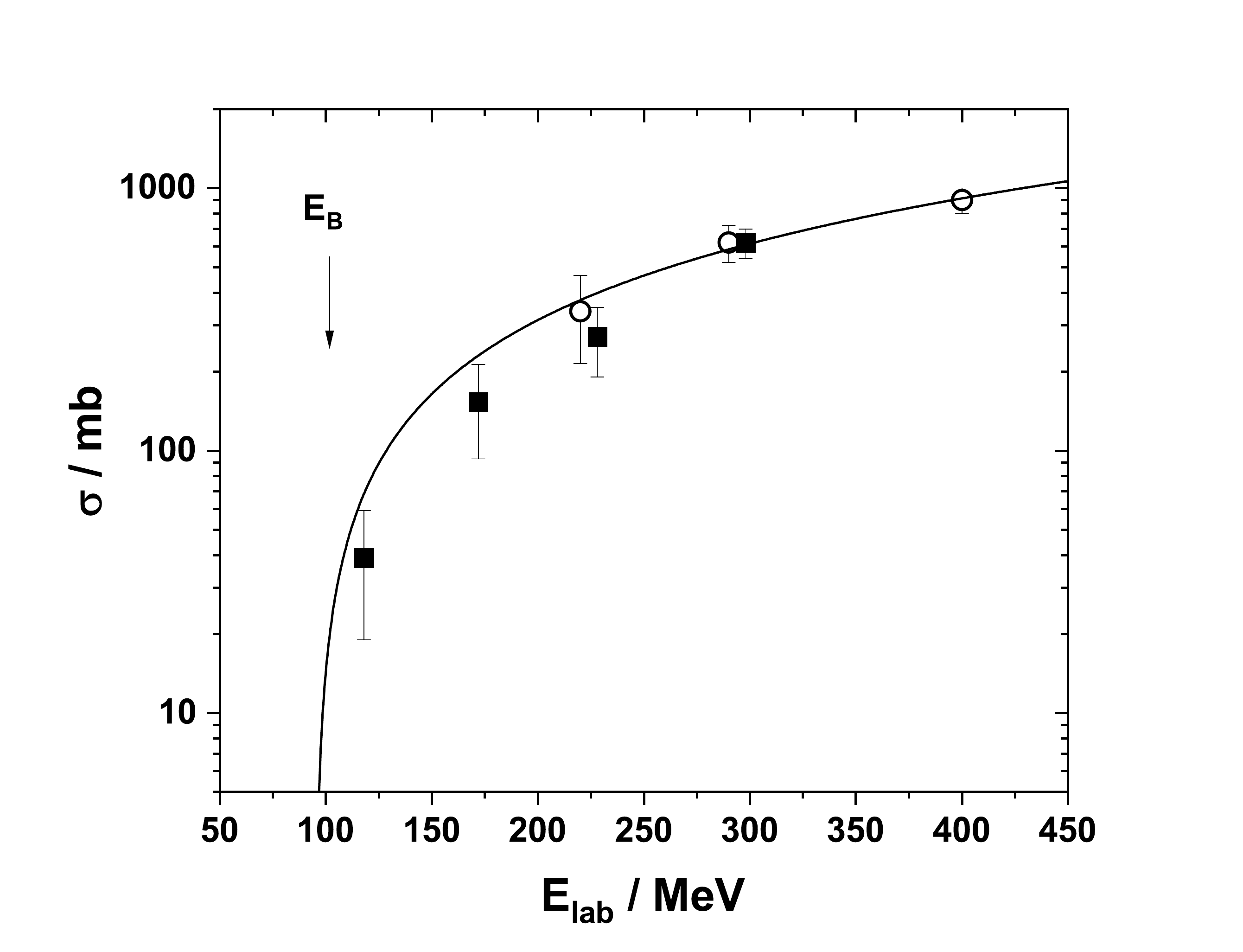}
	}
	\caption{Intensities of preequilibrium $\alpha$ - particles as a function of the projectile energy E$_{p}$. Open circles: data from \cite{Fuchs85};
		full dots: data from this work, normalised to those from \cite{Fuchs85}. The full line is the result of fitting a straight line the data of 
		\cite{Fuchs85}.
	}
	\label{fig:10}       
\end{figure*}

\vspace{8mm}
\subsection{\bf{5.3 Velocity Distributions}}
Here we again want to mention the velocity distributions and give one more comment. In fig. 9 we present a comparison of mean velocities for residues from Ar - induced reactions
at E = 14.5 AMeV (full symbols) \cite{Hess94a} and Ne - induced reactions  (open symbols); the circles refer to data from this experiment for $^{20}$Ne + $^{197}$Au,
the hexagon for $^{170}$Os from $^{20}$Ne + $^{165}$Ho, the squares to data from $^{20}$Ne + $^{208}$Pb \cite{Hess94} at 15.0 AMeV. Both sets of data refer to similar projectile velocities above the Coulomb barrier.\\
For A$_{res}$ $<$ 200 there is some indication, from the different mean velocities of the residues from the $^{40}$Ar - induced reactions and  $^{170}$Os from $^{20}$Ne + $^{165}$Ho that the sum of the masses of the emitted preequilibrium particles  on the average is constant for the same projectile velocities irrespective of their masses. Better experimental data to support this evidence are necessary, since such a behavior would be in contradiction to the conclusions of several authors (see e.g. \cite{Greg86}) who stress out a dependence of the linear momentum transfer on the projectile velocity, or more precise, the relative velocity above the fusion barrier.\\
For A$_{res}$ $>$ 200 we observe a significant dependence of the mean velocity on both the residue masses and the masses of the compound nuclei. While the mass dependence of the mean velocity for a fixed compound nucleus reflects the influence of the preequilibrium emission on the mass of the residues (see sect. 5.1), the dependence on the masses of the compound nuclei reflects the high fission competition, due to which preequilibrium emission enhances the survival probability.\\
From this side differences in the mean velocities are somewhat 'arificial' and do not directly reflect the mean momentum transfer in the nuclear reaction, since the velocity distribution of the evaporation residues is further influenced by the deexcitation process. This is also visible for the reactions using different projectiles The residues from $^{40}$Ar + $^{181}$Ta $\rightarrow$ $^{221}$Pa ((E$^{*}$\,=\,371 MeV), full triangles) appear to have higher mean velocities than those from $^{20}$Ne + $^{197}$Au $\rightarrow$ $^{217}$Ac (E$^{*}$\,=\,227 MeV), although the masses of the compound nuclei are similar, while the excitation energies are significantly different. So the residue velocity is not only determimed by the probability and multiplicity for preequilibrium emission but also from the number of thermal particles emitted before fission can compete with particles emission, because the mass and the atomic number of this intermediate nucleus  will determine essentially the fission to evaporation ratio. By this measured velocity distributions of residues reflect processes leading to their production, but are less suited to draw 'global' conclusions on general aspects of complete or incomplete fusion.\\ 

\subsection{\bf{5.4 Preequilibrium $\alpha$ - particles}}
The intensities of the preequilibrium $\alpha$ - particles observed in our experiment are shown in fig. 10 (full dots) as a function of the projectile energy and are 
compared with the results of Fuchs et al. \cite{Fuchs85} (open circles). Normalising our data at E$_{p}$\,=\,298 MeV, to the result of \cite{Fuchs85} at E$_{p}$\,=\,290 MeV, we find good agreement 
also at E$_{p}$\,=\,228 MeV with the value reported in \cite{Fuchs85} at E$_{p}$\,=\,220 MeV. At the lower energies our data are principally in-line with a 
'smooth' trend from the values at higher energies (full line). This trend does not indicate a threshold energy for preequilibrium emission, but the steep decrease towards
the lowest energie rather seem to be connected with the sharp drop of the fusion cross section when the fusion barrier (E$_{B}$\,=\,101.7 MeV \cite{Bass74}) is reached.

\section{6. Conclusion}
Velocity and mass distributions of residues from bombardments of $^{197}$Au and $^{165}$Ho with $^{20}$Ne at E = 298 MeV have been investigated. It has been shown, that the residue production is predominantly accompanied by emission of preequilibrium particles, which results in mean velocities of the residues significantly lower than v$_{CN}$. Comparisons of the mass distributions of astatine and polonium isotopes from the irradiations of $^{197}$Au with results from evaporation codes  indicate a shift of the experimental distributions to higher mass values compared to the model calculations. The latter also indicate, that this shift can also not be explained by fusion after projectile break-up  according to the sum-rule model as it can at lower energies, so we tentatively assign it to preequilibrium emission from the interaction zone at the beginning of equilibration. But we have to note, that the production rates appear essentially higher for the heaviest residues than the prediction of the standard hybrid model. As those nuclei exhibit the lowest mean velocities, it seems likely that the emission rate of preequilibrium $\alpha$ particles is underpredicted.
We also measured light particle (protons, $\alpha$ particles, Li- and Be- nuclei) energy spectra. A more detailed analysis of the 
$\alpha$ particle - spectra was performed. Two components, one ascribed to 'thermal - evaporation' particles and one, having same velocities as the projectiles
to 'preequilibrium' particles. The intensity of the latter drops drastically towards lower bombarding enery, but there seems real threshold energy for
preequilibrium $\alpha$ - particle emission, it rather seems to vanish at E$_{lab}$ $\approx$ 100 MeV (E$^{*}$ $\approx$ 45 MeV) i.e. at energies around the fusion barrier.\\

\section{Acknowledgement}
The experiment was performed in collaboration with A. L\"uttgen (deceased) and V. Ninov in June, 1991.
The author thanks A. Sitarcik for providing an updated version of fig. 5.

%
%

\end{document}